\documentclass[a4paper]{jpconf}
\usepackage{graphicx}
\usepackage{amsmath}
\usepackage{amssymb}
\usepackage{siunitx}
\def \Lu{$^{176}$Lu$^+$}

\begin{document}
\title{Validating a lutetium frequency reference.}

\author{K J Arnold\textsuperscript{1,2}, S Bustabad\textsuperscript{3}, Zhao Qi\textsuperscript{1}, Qin Qichen\textsuperscript{1}, Zhiqiang Zhang\textsuperscript{1}, Zhang Zhao\textsuperscript{1} and  M D Barrett\textsuperscript{1,3}}

\address{\textsuperscript{1}Center for Quantum Technologies, National University of Singapore, Singapore}
\address{\textsuperscript{2}Temasek Laboratories, National University of Singapore, Singapore}
\address{\textsuperscript{3}National Metrology Center, Agency for Science, Technology and Research (A*STAR), Singapore}

\ead{phybmd@nus.edu.sg}

\begin{abstract}
We review our progress in developing a frequency reference with singly ionized lutetium and give estimates of the levels of inaccuracy we expect to achieve in the near future with both the $^1S_0\leftrightarrow{}^3D_1$ and $^1S_0\leftrightarrow{}^3D_2$ transitions.  Based on established experimental results, we show that inaccuracies at the low $10^{-19}$ level are readily achievable for the $^1S_0\leftrightarrow{}^3D_1$ transition, and the frequency ratio between the two transitions is limited almost entirely by the BBR shift.  We argue that the frequency ratio measured within the one apparatus provides a well-defined metric to compare and establish the performance of remotely located systems.  For the measurement of an in situ frequency ratio, relativistic shifts drop out and both transitions experience the same electromagnetic environment.  Consequently, the uncertainty budget for the ratio is practically identical to the uncertainty budgets for the individual transitions.  If the ratios for two or more systems disagree we can be certain at least one of the clock assessments is incorrect.  If they agree, subsequent comparisons on one transition would only differ by relativistic effects.  Since motional effects are easily assessed and typically small for a heavy ion, only the differential gravitational red-shift will significantly contribute and this can be confirmed by comparison on the second transition.
\end{abstract}

\section{Introduction}
Over several years, we have developed the tools and methodologies to utilize singly-ionized lutetium (\Lu) as a frequency standard of the highest calibre.  Recently we have experimentally demonstrated agreement between two independent \Lu references at the low $10^{-18}$ level with clear evidence that mid $10^{-19}$ is readily attainable \cite{zhiqiang2023176lu}.  This has been achieved with much less technical effort than needed in other systems, as the clock transition used has the least sensitivity to environmental factors than any other system that we are aware of.  We would argue that many systems would be hard pressed to demonstrate the levels of accuracy that lutetium so easily affords.  

The relative ease at which a comparison at the $10^{-18}$ level can be achieved with lutetium should not be under-rated.  Comparisons between clock systems claiming this level of accuracy have variations of measured ratios that cannot be explained by uncertainty budgets of the individual systems, which is compelling evidence that large systematic shifts are not being controlled to the levels claimed.  Such problems are unlikely to occur for lutetium for the simple fact that systematics are not large enough to cause such variations.  Indeed, the total shift and any single contribution to the uncertainty budget given in \cite{zhiqiang2023176lu} is, at most, at the level of a few $10^{-18}$.  Moreover, it is self-evident that such low systematics will have substantial room for improvement if current state-of-the-art technologies were employed to take advantage of this potential. 

At the level of $10^{-18}$, an important question to ask is: how does one know that a clock is truly performing at the claimed level without a direct comparison.  Even if a comparison is made via an optical fiber link, the comparison would be influenced by the local gravitational potential, which is the basis for using optical clocks for geodesy \cite{mehlstaubler2018atomic,mcgrew2018atomic}.  This is a catch-22 situation in which a comparison to determine the differential gravitational potential would require a comparison using knowledge of the gravitational potential to ensure the clocks are sufficiently accurate to make the measurement.  Lutetium offers a unique solution to this problem in that it provides two high quality clock transitions within the one atom.  Thus a frequency ratio for each individual system can be measured and globally compared.  When measured on the same atom, a frequency ratio is independent of any relativistic effects, namely the gravitational redshift and motion.  Motional effects are easily assessed and typically small for a heavy ion, so the measured IFR provides independent confirmation on the performance of two remotely located systems: if the ratios for each system disagree, there must be something wrong with one of the assessments; conversely the ratios can only agree if the individual assessments are correct.  We neglect the very unlikely scenario that two independently assessed systematics of completely different physical origin are wrong and just so happen to cancel each other out, particularly given the low absolute systematic shifts that lutetium affords. No other atomic system that we are aware of could provide such a validation at least not to the degree that is possible with \Lu.

\section{Systematics}
In this section, we discuss the leading systematics for \Lu and the level of assessment that can be made for both $^1S_0\leftrightarrow{}^3D_1$ and $^1S_0\leftrightarrow{}^3D_2$ transitions, which we henceforth refer to as the primary and secondary transitions respectively.  In both cases we assume that hyperfine averaging (HA) has been used to suppress shifts associated with the electronic angular momentum as first proposed in \cite{barrett2015developing}.  For the primary transition it is also assumed hyper-Ramsey spectroscopy is used to suppress the probe-induced ac-Stark shift.  

A summary of typical systematics is given in table~\ref{Systematics}.  For the most part, values given in the table are based on measurements that have been carried out in the lab.  The two most notable exceptions are the residual quadrupole shift for the secondary transition and the blackbody radiation shift  as discussed in sections~\ref{Systematic:RQS} and~\ref{Systematic:BBR} respectively.  

\begin{table}[h]
\caption{\label{Systematics} Expected uncertainty budgets for upcoming comparisons.  All values are expressed fractionally in units of $10^{-18}$. Shifts are typical values measured in our current systems.  Uncertainties are based on those reported in \cite{zhiqiang2023176lu} along with expected improvements as discussed in the text. Uncertainties are added in quadrature, but for the secondary transition ($^1S_0\leftrightarrow{}^3D_2$)  we omit the BBR shift to better illustrate the extent to which it dominates the uncertainty budget.}
\begin{center}
\lineup
\begin{tabular}{l S[table-format=2.2] S[table-format=2.2] S[table-format=2.2] S[table-format=2.2]}
\br
& \multicolumn{2}{c}{$^3D_1$} &  \multicolumn{2}{c}{$^3D_2$} \\
Effect & \multicolumn{1}{c}{Shift} & \multicolumn{1}{c}{Unc.} & \multicolumn{1}{c}{Shift} & \multicolumn{1}{c}{Unc.} \\                   
\mr
Quad. Zeeman ($0.1\,\mathrm{mT}$) & {$-$138} & 0.04 & {\048.1} &0.04 \\
ac-Zeeman (rf) & 0.54 & 0.01 & {$-$10.5} & 0.11 \\
ac-Zeeman (microwave) & -0.10 & 0.05 & 0.00 & 0.00 \\
2\textsuperscript{nd}-order Doppler (thermal) & -0.13 & 0.06 & -0.13 & 0.06 \\
Micromotion &-0.10 & 0.10 & -0.10 & 0.10 \\
RQS & 0.22 & 0.02 & \0{$\lesssim 0.5$} & 0.10 \\
Microwave coupling & 0.00 & 0.21 & 0.00 &0.21 \\
BBR $(300\pm 5\,\mathrm{K})$ & -1.36 & 0.16 & {\027.0} & {\01.8} \\
\mr
\textbf{Total} & & 0.30 & & 0.28 \\
\br
\end{tabular}
\end{center}
\end{table}

\subsection{Quadratic Zeeman shifts}
\label{Sect:QZS}
For a given transition, quadratic Zeeman shifts are given by 
\begin{equation}
\label{Eq:QZS}
\alpha_z \left(B_0^2 + \langle \tilde{B}^2 \rangle\right)+\tilde{\alpha}_z \langle \tilde{B}_\perp^2 \rangle,
\end{equation}
where $B_0$ is the applied static field, $\tilde{B}$ is the amplitude of an oscillating field, and $\tilde{B}_\perp$ is the component of $\tilde{B}$ perpendicular the $B_0$.  For lutetium, the most significant oscillating field is from radio-frequency (rf) currents in the electrodes driven by the rf trapping potential.  The coefficient $\alpha_z$ arises from magnetic coupling to other fine-structure manifolds and $\tilde{\alpha}_z$ arises from magnetic coupling between hyperfine levels of the fine-structure manifold of interest \cite{gan2018oscillating}.  For the primary transition, the measured value of $\alpha_z$ is $-4.89264(88)\,\mathrm{Hz/mT^2}$ \cite{zhiqiang2023176lu}, which gives a fractional frequency shift of $\sim 1.38\times 10^{-16}$ at our nominal operating field of $B_0\sim0.1\,\mathrm{mT}$.  Since $B_0$ is tracked during clock operation at the level of $10\,\mathrm{nT}$, the static term only affects an uncertainty budget at the mid $10^{-20}$ level.  

The value of $\tilde{\alpha}_z$ is calculated to give a fractional shift of $5.65\times 10^{-19}$ for an rms amplitude of $1\,\mu\mathrm{T}$,  which we take to be accurate at the $1\%$ level\footnote{Magnetic field effects depend mostly on angular momentum considerations and are typically accurately calculated.}.  Indeed, calculation of the quadratic Zeeman shifts of individual hyperfine splittings differ only by Clebsch-Gordan factors and agree with measured values to better than $0.5\%$.  As proposed in \cite{gan2018oscillating} and demonstrated in \cite{arnold2020precision}, the amplitude of $\tilde{B}_\perp$ is measured via an Autler-Townes splitting with an inaccuracy better than 0.1\% and different orientations of $B_0$ can be used to infer $\tilde{B}$ \cite{kaewuam2020precision}.  In our current traps $\tilde{B}\lesssim 1\mu\mathrm{T}$, which keeps the contributions to an uncertainty budget below $10^{-20}$.  The uncertainty value given in the table has been rounded up to avoid a zero entry for a non-zero shift.

For the secondary transition, $\alpha_z$ is yet to be measured, but is calculated to give a fractional shift of  $4.8\times 10^{-17}$ at our nominal value of $B_0$.  We expect this to be measurable to the same level as the primary transition and to similarly contribute to an uncertainty budget at the low $10^{-20}$ level.  However, the value of $\tilde{\alpha}_z$ is a factor of 20 larger giving a calculated fractional frequency shift of  $-1.05\times 10^{-17}$ for an rms amplitude of $1\,\mu\mathrm{T}$ for $\tilde{B}$.  Consequently, this term will limit the Zeeman uncertainties to $1.1\times10^{-19}$.  In principle, $\tilde{\alpha}_z$ could be assessed more accurately, but the shift can also be suppressed by lowering the trap drive frequency, $\Omega_\mathrm{rf}$.  For a given $\Omega_\mathrm{rf}$ and voltage amplitude $U_0$, the trap psuedo-potential confinement frequencies, $\omega_i$, scale as $U_0/\Omega_\mathrm{rf}$ whereas the rf currents scale as $U_0\Omega_\mathrm{rf}$.  Consequently, for fixed $\omega_i$, $\langle \tilde{B}^2 \rangle$ and $\langle \tilde{B}_\perp^2 \rangle$ scale as $\Omega_\mathrm{rf}^4$.

It is worthwhile commenting on the stability of the $\tilde{B}_\perp$ as comments in the literature are overly pessimistic \cite{beloy2023trap} and misrepresent our own results \cite{arnold2020precision}.  The magnitude of $\tilde{B}_\perp$ is directly proportional $\omega_i$ and the orientation is fixed by the trap geometry however imperfect that might be.  The trap geometry certainly doesn't change, and even a $1\%$ change in $\omega_i$ is a substantial change in trap frequencies.  It is also a significant change in the orientation of $B_0$, which influences optical pumping and state preparation in our case.  Indeed, in our experiments, the orientation of $B_0$ is set relative to the linear polarization of the optical pumping beam that prepares the atom in $|{}^3D_1,F=7,m_F=0\rangle$.  Maximizing the depumping time out $|{}^3D_1,F=7,m_F=0\rangle$ enables alignment of $B_0$ to better than half a degree.  One can expect some level of inhomogeneity in the spatial dependence of $\tilde{B}_\perp$, but not significantly over distances bounded by micromotion compensation.  Consequently, there is no practical need to remeasure an Autler-Townes splitting on a regular basis as suggested in \cite{beloy2023trap}.  Only if there is a deliberate change in the pseudo-potential frequencies $\omega_i$ or to the clean-up polarizer in the optical pumping beam line would such a remeasurement be needed.  Of course, the situation would be very different if the rf-currents and associated shifts were substantially larger than we see in our current traps.

For the primary transition there is an additional contribution to the ac-Zeeman shift that arises from our interrogation technique, which utilizes microwave transitions between the upper clock states to effect an averaging over all $m=0$ hyperfine states.  The technique, which we refer to as hyperfine-averaged Ramsey spectroscopy (HARS), was first demonstrated in \cite{kaewuam2020hyperfine}, and a more detailed analysis given in \cite[Supplemental]{zhiqiang2023176lu}.  The ac-Zeeman shifts arise from unwanted polarizations in the applied microwave fields that off-resonantly couple to neighbouring Zeeman states.  Polarization components can be measured directly by microwave spectroscopy and the most significant contributions to the ac-Zeeman shift arise from coupling to the $F=7$ states due to the very small $g_F$-factor.  Consequently, this is unlikely to have any significant influence for the secondary transition, for which $g_F$-factors are more than an order of magnitude larger.

\subsection{Micromotion}
Considerations of excess micromotion (EMM) are much the same for lutetium as any other ion.  For lutetium, we probe micromotion using sideband spectroscopy on the secondary clock transition, specifically the $\left|{}^1S_0,F=7,m_F=0\right\rangle \leftrightarrow \left|{}^3D_2,F=9,m_F=0\right\rangle$ transition, which does have benefits.  When using sideband spectroscopy, it is important to have good resolution of both the carrier and sideband, and to ensure one remains on resonance with the transition.  Exact numbers depend on specific beam orientations and polarizations relative to the applied magnetic field, but generally speaking it is straightforward to achieve a Rabi frequency of $\sim 10\,\mathrm{kHz}$ on the carrier with $<100\,\mathrm{\mu W}$ of optical power, which is readily available at the clock wavelength of $804\,\mathrm{nm}$.  The 10\,kHz Rabi frequency is well below any trap frequencies and has an associated ac-Stark shift $\lesssim 10\,\mathrm{Hz}$.  This readily allows the detection of micromotion sidebands corresponding to a modulation index below 0.01.  For typical trap drive frequencies of $\Omega_\mathrm{rf}=2\pi\times 20\,\mathrm{MHz}$, the associated fractional second-order Doppler shift (SDS) is below $10^{-19}$.  

The fractional shift from EMM also has an ac-Stark shift arising from the electric fields driving the motion.  For the primary transition this contribution is practically negligible owing to the small value of $\Delta\alpha_0(0)$.  For the secondary transition, the contribution has opposite sign to the SDS leading to a cancellation as is well known for clock transitions having $\Delta\alpha_0(0)<0$.  This leads to the concept of a ``magic rf'' at which EMM shifts cancel to leading order.  For the secondary transition in \Lu, this occurs at approximately $\Omega_\mathrm{rf}=2\pi\times 32\,\mathrm{MHz}$.

Micromotion shifts reported in \cite{zhiqiang2023176lu} were specified to bound the largest value observed over a period of several months with a timescale between measurements that was sometimes much longer than the measurement campaign.  We are now working towards a more autonomous operation with automated assessment and compensation of micromotion.  We consequently expect a significant reduction in this contribution to an uncertainty budget.  Preliminary work indicates shifts can be maintained below $10^{-19}$ with compensation carried out once or twice a day with minimal impact on run time.

\subsection{Thermal motion}
Thermal motion is much less problematic for heavy ions.  The cooling transition in \Lu has a linewidth of $\Gamma=2\pi\times 2.45\,\mathrm{MHz}$, which leads to a fractional SDS at the Doppler cooling limit of $-7.7\times 10^{-20}$ including the contribution from intrinsic micromotion (IMM) \cite{berkeland1998minimization}.  We typically get within about a factor of two of this limit.  As this is a relativistic shift, it is the same for both transitions and cancels for an IFR measurement.  In our recent comparison \cite{zhiqiang2023176lu}, the shift for one trap is substantially larger owing to a significant heating problem of unknown origin.  The trap has now been replaced and gives a heating rate consistent with that found in our second trap.  We currently use dephasing measurements of the clock transition to infer the thermal state, which is sufficient for any levels of precision we could achieve in the near future.

As shown in \cite{berkeland1998minimization}, there is an additional frequency shift from thermal motion due to sampling of the electric fields over the spatial extent of the external motion.  As for EMM this is negligibly small for the primary transition owing to the small value of $\Delta\alpha_0(0)$ and leads to some level of cancellation for the secondary transition owing to the sign of $\Delta\alpha_0(0)$.  The uncertainty given in the table for $^3D_2$ is therefore a conservative estimate inferred from measurements on $^3D_1$.

\subsection{Residual quadrupole shift}
\label{Systematic:RQS}
Mixing of fine-structure levels due to the hyperfine interaction brings about modifications to Land\'{e} $g$-factors and quadrupole moments for a given state \cite{zhiqiang2020hyperfine,beloy2017hyperfine}.  These corrections are not cancelled by HA giving rise to a residual quadrupole moment (RQM) for the hyperfine-averaged frequency reference.  For lutetium, leading order corrections for both $g$-factors and quadrupole moments arise from the magnetic dipole moment of the nucleus.  Thus, high accuracy $g$-factors measurements can be used to factor out matrix elements associated with the nuclear magnetic dipole moment for the estimation of the RQM.  Using this approach, we have obtained an estimate of $-2.54\times 10^{-4}\,ea_0^2$ for the RQM of $^3D_1$ \cite{zhiqiang2020hyperfine}, which leads to a shift at low $10^{-19}$ for typical confinement conditions.  A measurement of the shift can then be obtained by scaling the quadrupole shift extracted from clock operation with Ba$^+$ by the ratio of quadrupole moments.  An estimate of the RQM for $^3D_2$ could be achieved in a similar fashion.  However, contributions to the $^3D_2$ RQM are primarily from coupling to $^3D_1$ and $^3D_3$.  The former can be inferred from the RQM of $^3D_1$ and the latter contribution is reduced by the larger energy splitting.  Consequently, we expect the RQM for $^3D_2$ to have a similar magnitude to $^3D_1$ barring any fortuitous cancellation. 

A more direct experimental assessment of the RQM could potentially be obtained from differential quadrupole shifts of hyperfine splittings in a multi-ion crystal as used to determine the quadrupole moments of $^3D_1$ and $^3D_2$ in \cite{kaewuam2020precision}.  The quadrupole moments were extracted from measurements of a single hyperfine splitting.  To lowest order, ratios of the quadrupole shifts for different hyperfine splittings are completely determined by Clebsch-Gordan coefficients.  However, corrections to these ratios from mixing with other fine-structure levels can also be related to the RQM.  Indeed it was noted in \cite{kaewuam2020precision} that the inferred quadrupole moments may well be influenced by this.

\subsection{Probe-induced ac-Stark shift}
The lifetime of the primary clock transition is estimated to be about 1 week, and this gives rise to a significant probe-induced ac Stark shift, although it is substantially smaller than for the E3 transition in Yb$^+$ under similar interrogation conditions.  For the experiments reported in \cite{zhiqiang2023176lu}, Ramsey spectroscopy with an interrogation time of 744\,ms and optical pulses with an 8\,ms $\pi$-pulse time, gave a shift of $\sim 10^{-16}$ with an uncertainty at mid $10^{-18}$ level.  Further use of hyper-Ramsey spectroscopy practically eliminates the shift without any need to monitor or control variations in the laser intensity beyond the usual power stabilization.  For the secondary transition, the lifetime is substantially reduced and the probe-induced ac-Stark shift is negligible under any reasonable interrogation conditions.

The probe-induced ac-Stark shift is dependent on the interrogation time and $\pi$-pulse duration.  The ac-Stark shift itself scales as $\tau_L^{-2}$, where $\tau_L$ is the $\pi$-pulse time, and Ramsey spectroscopy gives a suppression factor that scales as $T/\tau_L$, where $T$ is the Ramsey time. Consequently, the probe-induced ac-Stark shift becomes much less problematic as laser technology allows increasing interrogation times.  For example, increasing the interrogation time to 10\,s and $\tau_L$ to 24\,ms, reduces the shift for the primary clock transition to $5\times 10^{-18}$ with an uncertainty of $2.5\times 10^{-19}$ assuming a modest $5\%$ stability of the laser intensity.  At this level, hyper-Ramsey spectroscopy may well be unnecessary.

\subsection{Blackbody radiation shift}
\label{Systematic:BBR}
The blackbody radiation (BBR) shift is characterized by the differential static scalar polarizability, $\Delta\alpha_0(0)$, which has been measured for both transitions \cite{arnold2018blackbody} using a CO$_2$ laser.   For the primary transition, the shift is just $-1.36(9)\times 10^{-18}$ at 300\,K.  In our current experiments we do not have the ability to measure temperature but even an assessment of $35(10)\,\mathrm{{}^\circ C}$ would give a contribution of just $2.9\times 10^{-19}$ to an uncertainty budget.  A trap that incorporates a temperature sensor at the trap electrodes is currently under construction, but comparisons under any practical circumstances are unlikely to be influenced by temperature.  In contrast, the BBR shift for the secondary transition is about a factor of 20 larger with a fractional shift of $2.70(21)\times 10^{-17}$ at 300\,K.  Thus, inaccuracy below $10^{-18}$ on the secondary transition requires a significant improvement in the measured polarizability and a temperature assessment better than $\pm 3\,\mathrm{^\circ C}$.  

To improve the assessment of $\Delta\alpha_0(0)$ for the secondary clock transition, we plan to make a comparison measurement using Ba$^+$ as proposed in \cite{barrett2019polarizability}.  The basic idea is to use the more accurate value of $\Delta\alpha_0(\omega)$ for Ba$^+$ to better characterize the in situ intensity of the CO$_2$ laser.  Supporting measurements to characterize $\Delta\alpha_0(\omega)$ for Ba$^+$  have already been carried out \cite{arnold2019measurements, zhang2020branching, woods2010dipole} that enable an improvement in the fractional inaccuracy of $\Delta\alpha_0(0)$ to $0.5\%$ \cite{chanu2020magic}.  This will reduce the uncertainty contribution from $\Delta\alpha_0(0)$ to $< 2\times 10^{-19}$ for operation at 300\,K leaving temperature assessment as the biggest hurdle for the secondary transition.

As $\Delta\alpha_0(\omega)<0$ for the secondary transition, one could in principle use the approach used for Sr$^+$ and Ca$^+$ \cite{dube2014high,huang2019ca+}, which makes use of the magic rf for the trap drive at which large EMM shifts are cancelled.  However, both EMM shifts and the quadratic Zeeman shift from rf currents in the electrodes vary quadratically as the ion is displaced.  With very large amounts of EMM, the contribution from the quadratic Zeeman shift can be expected to be prohibitively large.  Moreover, we measure this with Ba$^+$ for which uncompensated micromotion will lead to a different spatial location than Lu$^+$ owing to the difference in mass.   

\subsection{Miscellaneous}
\label{Misc}
Additional shifts that have been considered but outside the categories given above are phase chirp and uncertainties arising from the imperfections in the HARS interrogation technique we have used.  Phase chirp has been investigated following the methodology reported in \cite{falke2012delivering} and estimated to be less than $10^{-19}$.  When using HARS, the most important effects are ac-Zeeman shifts from the microwave fields as discussed in Sect.~\ref{Sect:QZS}, and from the accuracy of the microwave $\pi$-pulse times, which is the more significant issue.  We estimate the sensitivity to the $\pi$-pulse times from simulations, and we expect this to be similar for both transitions.

At the level of $\lesssim 10^{-19}$ there may well be additional shifts, such as collision shifts that appear in some uncertainty budgets, to consider.  Without the measurement precision or methodology to investigate further, we consider such considerations to be academic and speculative.  Until measurement precision reaches the level of realistic estimates, we omit them.

\section{An in situ frequency ratio measurement}
On the primary clock transition, inaccuracy below $5\times 10^{-19}$ is readily achievable in the near future.  Eliminating the large heating rate of one trap, characterizing the quadrupole fields as already done for one trap, and automated micromotion detection and compensation reduces the largest contributions to the uncertainty budgets given in \cite{zhiqiang2023176lu} to $\lesssim 10^{-19}$.  With a realistically achievable temperature assessment with an inaccuracy of $\pm 5\,\mathrm{^\circ C}$, the total uncertainty would be $3.0\times 10^{-19}$, with a significant contribution coming from imperfections in the interrogation technique.  The only shift that could realistically affect the transition at that level would be that from the dc applied magnetic field, which is monitored with an interleaved servo.  With a 10\,s long interrogation via correlation spectroscopy, verification at the level of $5\times 10^{-19}$ would be achievable with a few days of integration.  

The BBR shift not withstanding, the same argument applies to the secondary transition.  Consequently, measurement  of an in situ frequency ratio (IFR) would be almost entirely limited by the BBR shift on the secondary transition as is evident from table~\ref{Systematics}.  At a liquid nitrogen temperature of 77\,K, the BBR shift is reduced to $10^{-19}$ and removes this limitation.  The measured IFR then becomes an independent figure of merit for clock performance at the level of uncertainty producible by either transition.  At room temperature, a $5\times 10^{-19}$ uncertainty would require a temperature assessment with an uncertainty $\approx \pm 1.5\,\mathrm{^\circ C}$, which is fairly challenging in practice.  Arguably, so is the use of cryogenic temperatures, but a measurement of the IFR at 77\,K is practically a zero temperature measurement.  The corresponding ratio is a property of the atom that can be measured independently by any lab and this can achieved with a given uncertainty if and only if both clock transitions can operate at the same level of performance, which is currently at the level of $\lesssim 5 \times 10^{-19}$.  

Once a zero temperature ratio is established, it would allow an IFR measured at room temperature to estimate the BBR shift and hence the effective temperature of the thermal environment seen by the ion.  If we assume both the zero-temperature and room-temperature ratios were measured with an inaccuracy of $5\times 10^{-19}$, the contribution from the BBR shift for the room temperature system could be inferred with an inaccuracy of $5\sqrt{2}\times 10^{-19}$ corresponding to a temperature inaccuracy of $\pm 2^\circ \mathrm{C}$.  Two clocks having been assessed in this manner could then be compared on the primary transition to determine the gravitational red-shift between them, with a small correction due to the temperature.  Subsequent comparison on the secondary transition could then be used as a consistency check to confirm the gravitational red-shift and temperature assessment.

Even without the determination of a zero-temperature frequency ratio, use of the secondary transition would still be instrumental in bounding the temperature difference between two clocks.  For the secondary transition, the BBR shift is the only significant systematic above mid $10^{-19}$, so any difference between two systems would be attributable to temperature.  For a given difference in the secondary frequency or primary to secondary ratio, the temperature difference is a monotonically decreasing function of the mean temperature.  Taking the temperature of the surrounding as a lower bound for the temperature of the coldest clock, one can then infer bounds on the possible temperature difference between the two systems.  

\section{Stability Considerations}
The level of accuracy attainable with lutetium requires an improvement in stability beyond that achievable with our available laser system.  For this reason we have relied upon correlation spectroscopy to extend the achievable interrogation time and validate the quality of a lutetium frequency reference.  Nevertheless, there is still a clear need to increase the number of ions beyond the single ion paradigm and proof-of-principle experiments were first demonstrated in \cite{tan2019suppressing, kaewuam2020hyperfine}.  In those experiments we demonstrated the ability to measure and control inhomogeneous shifts to a degree that would allow long (10\,s) interrogation times.  

For lutetium the most notable source of inhomogeneous broadening is the quadrupole shift induced by the electric fields from neighbouring ions.  As demonstrated in \cite{tan2019suppressing}, this can be mitigated by aligning the magnetic field at angle $\theta_0=\cos^{-1}(1/\sqrt{3})$ to the ion crystal at which the contribution from neighbouring ions is zero.  Although the sensitivity of the suppression is linear in the angle, this should be considered in the context of how well the angle can be set and maintained.  As demonstrated in \cite{kaewuam2020precision}, the differential shift between the outmost and innermost ions in a linear ion crystal can be measured using microwave spectroscopy.  In that work the field was aligned to maximize the shift and the measured value used to estimate the quadrupole moment.  Alternatively, the same method can be used to align the magnetic field to zero the difference.

For the purposes of illustration, consider a 10-ion crystal with an axial confinement of 100\,kHz.  For angles $\theta=\theta_0+\delta\theta$, the differential quadrupole shift between the outermost and innermost ions is approximately $7\delta\theta$\,Hz when interrogating the $F=7$ to $F=8$ microwave transition.  It is reasonably straightforward to keep magnetic field variations below $0.1\,\mathrm{\mu T}$ at our operating field of 0.1\,mT, which corresponds to $\delta\theta=10^{-3}$ and reduces the differential quadrupole shift to 7\,mHz.  Due to the low magnetic field sensitivity of the clock states and practically unlimited lifetime of $^3D_1$, Ramsey interrogation can easily resolve the 7\,mHz in a matter of minutes.  Moreover, any remaining inhomogeneity would be removed during interrogation when using HARS as demonstrated in \cite{kaewuam2020hyperfine}.

Chip scale ion traps may eventually make it feasible to consider hundreds or even thousands of ions, but near term advances will undoubtedly utilize a more modest number.  In any case, it will always be advantageous to extend interrogation times to the limit that technology will allow.  Already laser technology permits interrogation times of a few tens of seconds.  For the primary transition, a 5-ion lutetium clock with a 10\,s interrogation time would provide a projection noise limit of $6.4\times 10^{-17}/\sqrt{\tau}$, which allows mid $10^{-19}$ precision in 5 hours.  The corresponding instability for the secondary transition is slightly larger ($8.0\times 10^{-17}/\sqrt{\tau}$) owing to the finite lifetime, but this still enables mid $10^{-19}$ precision within 7 hours.   From the perspective of gaining stability, the importance of lifetime is best illustrated by considering an atom such as In$^+$.  With a lifetime of just 200\,ms, the same $6.4\times 10^{-17}/\sqrt{\tau}$ instability achievable with 5 Lu$^+$ ions would require over 50 In$^+$ ions.

Another approach to improving stability of ion-based clocks is the use of differential spectroscopy demonstrated in \cite{kim2023improved}. In essence, a more stable lattice clock was used to extend the interrogation time of the ion beyond the coherence time of the laser to improve the measurement precision of the frequency ratio.  The measured frequency ratio was not given in their report as neither clock was operating with previously claimed levels of performance.  Nevertheless the method is of interest for those ions that can provide and maintain superior accuracy.  The method equally applies to ensembles of ions and even entangled states.  Entanglement could only be advantageous for clock transitions in which lifetime does not become a factor.  Given the week-long lifetime of $^3D_1$ and low sensitivity of the primary clock transition to its electromagnetic environment, this approach is ideally suited to lutetium.  

\section{Discussion}
In \cite{dimarcq2023roadmap}, 37 authors from 21 institutes across 15 different countries outline a roadmap towards the redefinition of the second based on optical frequency standards (OFS).  The roadmap lists a set of mandatory criteria to be met for such a redefinition and gives the degree to which the 37 authors consider these criteria to be met.  As this is presumably a representation of what is considered to be state-of-the-art, it is of interest to consider our work in the context of these criteria and in comparison to those works considered to have met those criteria.  Criteria most relevant to our work are I.1, and I.2, which we give here for convenience.

Mandatory Criteria I.1.b: \emph{At least three frequency evaluations of optical frequency standards based on different reference transitions, either in the same institute or different institutes, have demonstrated evaluated uncertainties $\lesssim 2\times 10^{-18}$ based on comprehensive, comparable and published accuracy budgets}.  This criteria is considered to be more or less fulfilled with one of the three evaluations quoted being that of the NIST Al$^+$ clock at $9.4\times 10^{-19}$ \cite{brewer2019al+}.  In that assessment, the researchers observe a $10^{-16}$ frequency difference when the atom is probed from counter-propagating directions.  This is assumed to be a first order Doppler shift of $5\times 10^{-17}$ arising from charging and discharging induced by photo-electrons generated by the UV lasers used in the operation of the clock.  But no conclusive scientific evidence is given to support this conjecture and no clock comparison has ever been presented that demonstrates cancellation of the shift.

Mandatory Criteria I.2.a: \emph{Unit ratios (frequency comparison between standards with same clock transition): at least three measurements between OFS in different institutes in agreement with an overall uncertainty of the comparison $\Delta\nu/\nu \lesssim 5\times 10^{-18}$ (either by transportable clocks or advanced links).  Applicable to at least one radiation of I.1.}  This is closely related to Mandatory Criteria I.1.a with the PTB Yb$^+$ comparison \cite{sanner2019optical} cited as one of the three comparisons being a ``first step in the right direction''.  We caution that a frequency comparison in and of itself does not validate an uncertainty budget, particularly if every effort is made to make the environmental conditions the same for both systems.  If temperatures are made near equal, simulations used to decide absolute temperature and extrapolations to determine the differential scalar polarizability are inconsequential.  For the latter, it is readily shown that the accuracy of the extrapolation given in \cite{huntemann2016single} is incorrect by at least a factor of three.

So how does lutetium stack up against this road map and the work of leading metrology institutes?  For our comparison reported in \cite{zhiqiang2023176lu}, the uncertainly for each system is $\leq 0.68 \times 10^{-18}$, although we openly acknowledge that this does not include the BBR shift as we are unable to measure temperature in either system.  However, if we assume the temperature of each system is within the range of $25-65\,^{\circ}\mathrm{C}$, the total uncertainty would still be below the $0.94\times 10^{-18}$ claimed by NIST \cite{brewer2019al+}.  As assumptions go, we consider this temperature range to be very reasonable.  Moreover, a temperature range of $25-110\,^{\circ}\mathrm{C}$ ensures the uncertainty for each system is below the roadmaps $2\times 10^{-18}$ requirement.  Additionally, given the low systematics of a  \Lu\,primary, a deviation by more than $2\times 10^{-18}$ would be highly unlikely.  With regards to criteria I.2.a, the comparison in \cite{zhiqiang2023176lu} has a projection noise limit of $3\times 10^{-18}$, which is below the $5\times 10^{-18}$ of criteria I.2.a.  

But what of frequency ratios between standards?  Frequency ratios with other OFS are only meaningful if carried out with the highest degree of scientific integrity and that starts with each OFS being independently and rigorously assessed.  Such an assessment would at least include: (i) a clear set of experimental measurements of atomic properties and environmental factors determining the performance claim, (ii) a same-species comparison with a precision to at least the level of accuracy being claimed, which cannot be better than the measurement precision, and (iii) a stress test on any systematic that is more than an order of magnitude larger than the accuracy claim i.e. a frequency comparison in which the systematic is deliberately shifted to give a predicted shift in the clock frequency.  We see a stand-alone uncertainty budget without a same-species comparison as an hypothesis that is to be experimentally tested, which can only be done through a same-species comparison.

Problems that arise for frequency ratios with other OFS should be resolved using impartial, evidence-based investigation.  Dismissing inconsistent results with error scaling \cite{dorscher2021optical} or a ``comprehensive Baysian analysis'' \cite{collaboration2021frequency}, or disregarding earlier measurements in conflict with the new \cite{collaboration2021frequency} has no place in precision metrology.  It undermines the purpose of carrying out a frequency comparison in the first place; be it a search for new physics, geodesy, or ensuring the integrity of OFSs around the world.  With the widespread acceptance of error scaling practices, any evidence of a variation of fundamental constants or postulated interactions with dark matter will be based on a subjective decision of how or when error scaling should be applied.  Equally, it would make ``option 2'' for the redefinition of the second questionable as the choice of weights for the contributing OFSs would succumb to the same considerations.  

We have argued that the frequency ratio between the primary and secondary transitions in \Lu\, is an independent figure of merit that confirms the performance of the individual transitions to the level they have been assessed.  Our approach to validating a \Lu\, frequency reference will be to establish two liquid nitrogen systems; one at the Center for Quantum Technologies and one at the National Metrology Center in Singapore.  The aim is to verify agreement of the IFR for the two systems, and then compare the two systems on the primary to deduce the local gravitational redshift, which can be independently confirmed by conventional technologies and also by comparison on the secondary.  Further comparisons with a room temperature system can then confirm the IFR may be used to assess the system temperature.  We would then have a complete set of well-defined experimental metrics to characterize and assure the performance of the standard.  

\ack
This research is supported by the Agency for Science, Technology and Research (A*STAR) under Project No. C210917001; the National Research Foundation (NRF), Singapore, under its Quantum Engineering Programme (QEP-P5); the National Research Foundation, Singapore and A*STAR under its Quantum Engineering Programme (NRF2021-QEP2-01-P03) and its CQT Bridging Grant; and the Ministry of Education, Singapore under its Academic Research Fund Tier 2 (MOE-T2EP50120-0014).

\bibliography{9FSM2023}

\providecommand{\newblock}{}
\begin{thebibliography}{10}
\expandafter\ifx\csname url\endcsname\relax
  \def\url#1{{\tt #1}}\fi
\expandafter\ifx\csname urlprefix\endcsname\relax\def\urlprefix{URL }\fi
\providecommand{\eprint}[2][]{\url{#2}}

\bibitem{zhiqiang2023176lu}
Zhang Z, Arnold K~J, Kaewuam R and Barrett M~D 2023 {\em Science Advances\/}
  {\bf 9} eadg1971

\bibitem{mehlstaubler2018atomic}
Mehlst{\"a}ubler T~E, Grosche G, Lisdat C, Schmidt P~O and Denker H 2018 {\em
  Reports on Progress in Physics\/} {\bf 81} 064401

\bibitem{mcgrew2018atomic}
McGrew W, Zhang X, Fasano R, Sch{\"a}ffer S, Beloy K, Nicolodi D, Brown R,
  Hinkley N, Milani G, Schioppo M {\em et~al.\/} 2018 {\em Nature\/} {\bf 564}
  87--90

\bibitem{barrett2015developing}
Barrett M 2015 {\em New Journal of Physics\/} {\bf 17} 053024

\bibitem{gan2018oscillating}
Gan H~C~J, Maslennikov G, Tseng K~W, Tan T~R, Kaewuam R, Arnold K~J,
  Matsukevich D and Barrett M~D 2018 {\em Physical Review A\/} {\bf 98} 032514

\bibitem{arnold2020precision}
Arnold K~J, Kaewuam R, Chanu S~R, Tan T~R, Zhang Z and Barrett M~D 2020 {\em
  Physical Review Letters\/} {\bf 124} 193001

\bibitem{kaewuam2020precision}
Kaewuam R, Tan T~R, Zhang Z, Arnold K~J, Safronova M~S and Barrett M~D 2020
  {\em Physical Review A\/} {\bf 102} 042819

\bibitem{beloy2023trap}
Beloy K 2023 {\em Physical review letters\/} {\bf 130} 103201

\bibitem{kaewuam2020hyperfine}
Kaewuam R, Tan T~R, Arnold K~J, Chanu S~R, Zhang Z and Barrett M~D 2020 {\em
  Physical review letters\/} {\bf 124} 083202

\bibitem{berkeland1998minimization}
Berkeland D, Miller J, Bergquist J~C, Itano W~M and Wineland D~J 1998 {\em
  Journal of applied physics\/} {\bf 83} 5025--5033

\bibitem{zhiqiang2020hyperfine}
Zhang Z, Arnold K~J, Kaewuam R, Safronova M~S and Barrett M~D 2020 {\em
  Physical Review A\/} {\bf 102} 052834

\bibitem{beloy2017hyperfine}
Beloy K, Leibrandt D~R and Itano W~M 2017 {\em Physical Review A\/} {\bf 95}
  043405

\bibitem{arnold2018blackbody}
Arnold K~J, Kaewuam R, Roy A, Tan T~R and Barrett M~D 2018 {\em Nature
  communications\/} {\bf 9} 1650

\bibitem{barrett2019polarizability}
Barrett M, Arnold K and Safronova M 2019 {\em Physical Review A\/} {\bf 100}
  043418

\bibitem{arnold2019measurements}
Arnold K, Chanu S, Kaewuam R, Tan T, Yeo L, Zhang Z, Safronova M and Barrett M
  2019 {\em Physical Review A\/} {\bf 100} 032503

\bibitem{zhang2020branching}
Zhang Z, Arnold K, Chanu S, Kaewuam R, Safronova M and Barrett M 2020 {\em
  Physical Review A\/} {\bf 101} 062515

\bibitem{woods2010dipole}
Woods S~L, Hanni M, Lundeen S and Snow E~L 2010 {\em Physical Review A\/} {\bf
  82} 012506

\bibitem{chanu2020magic}
Chanu S, Koh V, Arnold K, Kaewuam R, Tan T, Zhang Z, Safronova M and Barrett M
  2020 {\em Physical Review A\/} {\bf 101} 042507

\bibitem{dube2014high}
Dub{\'e} P, Madej A~A, Tibbo M and Bernard J~E 2014 {\em Physical review
  letters\/} {\bf 112} 173002

\bibitem{huang2019ca+}
Huang Y, Guan H, Zeng M, Tang L and Gao K 2019 {\em Physical Review A\/} {\bf
  99} 011401

\bibitem{falke2012delivering}
Falke S, Misera M, Sterr U and Lisdat C 2012 {\em Applied Physics B\/} {\bf
  107} 301--311

\bibitem{tan2019suppressing}
Tan T~R, Kaewuam R, Arnold K~J, Chanu S~R, Zhang Z, Safronova M and Barrett M~D
  2019 {\em Phys. Rev. Lett.\/} {\bf 123} 063201

\bibitem{kim2023improved}
Kim M~E, McGrew W~F, Nardelli N~V, Clements E~R, Hassan Y~S, Zhang X, Valencia
  J~L, Leopardi H, Hume D~B, Fortier T~M {\em et~al.\/} 2023 {\em Nature
  Physics\/} {\bf 19} 25--29

\bibitem{dimarcq2023roadmap}
Dimarcq N, Gertsvolf M, Mileti G, Bize S, Oates C, Peik E, Calonico D, Ido T,
  Tavella P, Meynadier F {\em et~al.\/} 2023 {\em Metrologia\/}

\bibitem{brewer2019al+}
Brewer S~M, Chen J~S, Hankin A~M, Clements E~R, Chou C, Wineland D~J, Hume D~B
  and Leibrandt D~R 2019 {\em Physical review letters\/} {\bf 123} 033201

\bibitem{sanner2019optical}
Sanner C, Huntemann N, Lange R, Tamm C, Peik E, Safronova M~S and Porsev S~G
  2019 {\em Nature\/} {\bf 567} 204--208

\bibitem{huntemann2016single}
Huntemann N, Sanner C, Lipphardt B, Tamm C and Peik E 2016 {\em Physical review
  letters\/} {\bf 116} 063001

\bibitem{dorscher2021optical}
D{\"o}rscher S, Huntemann N, Schwarz R, Lange R, Benkler E, Lipphardt B, Sterr
  U, Peik E and Lisdat C 2021 {\em Metrologia\/} {\bf 58} 015005

\bibitem{collaboration2021frequency}
Beloy K {\em et~al.\/} 2021 {\em Nature\/} {\bf 591} 564--569

\end{thebibliography}
\end{document}